# Reduce, Reuse, Recycle: Building Greener Software

Kaushik Dutta, *University of South Florida, duttak@usf.edu*
Debra Vandermeer, *Florida International University, vanderd@fiu.edu*

**Abstract**

Technology use has grown rapidly in recent years. It is infused in virtually every aspect of organizational and individual life. This technology runs on servers, typically in data centers. As workloads grow, more serves are required. Each server incrementally adds to the energy consumption footprint of a data center. Currently, data centers account for more than one percent of all power usage worldwide. Clearly, energy efficiency is a significant concern for data centers. While many aspects of data center energy efficiency have received attention, energy consumption is rarely considered in software development organizations. In this work, we consider the energy consumption impacts of fundamental software operations, and demonstrate that non-trivial energy savings can be achieved in software by making energy-conscious decisions regarding basic aspects of programming. This work has significant potential for practical impact; applying the lessons learned in this study can lead to greener software.

## 1. Introduction and Related Work

Virtually every aspect of modern life is infused with technology. Organizations sell products and services via always-on e-commerce sites, and automate business processes internally. Cloud computing services, through usage-based pricing, have reduced the up-front capital investment costs involved in starting up an online business, allowing for the creation of new businesses that would otherwise have been cost-prohibitive to start. Individuals make extensive use of technology to support personal and professional lives through social networks, cloud-based services (e.g., collaboration tools, email, backup services), entertainment (e.g., streaming media, online multiplayer gaming), and a host of other services.

All these services run as software applications on servers, typically in data centers. As workloads grow, additional servers are added to handle the growth. Each incremental server adds to the power usage footprint of the data center, both in terms of the energy required to run the server itself, as well as the power required for cooling and auxiliary equipment.

Power usage in data centers is a significant concern. Koomey (2008) estimates that power usage by data centers worldwide doubled between 2000 and 2005, from 0.5% of all worldwide power usage in 2000 to 1% in 2005, with usage in the US accounting for roughly 0.4% of all worldwide power usage in 2005. This estimate predates several major online phenomena. To name a few: Facebook opened registrations to the general public in September 2006 (Arrington, 2006); Netflix started streaming movies in 2007 (Anderson, 2007); and the iPhone, the first smartphone, first went on sale in June 2007 (Ricknas, 2008). More recently, in May 2011, Amazon announced that e-books are outselling print books (Miller and Bosman, 2011). It is unlikely that the growth of application workloads has leveled off since 2005.

The issue of power use in data centers has been addressed along multiple fronts, from data center design to hardware design to virtualization. Data center managers have worked to improve the infrastructure within data centers to improve the efficiency of cooling systems and power delivery. Facebook has developed a simplified data center design that is 38% more efficient and 24% less expensive to build than benchmark state-of-the-art data centers, and has shared the specifications of the server and data center technology through the Open Compute Project (Open Compute Project, 2011). Google developed an innovative data center design for its new facility in Finland. The facility uses seawater from the Baltic Sea to provide chiller-less cooling, resulting in significant energy savings (Google, Inc., 2011).

Hardware designers are working on energy efficiency as well. For instance, processor designers have proposed chip designs that adjust power use based on workloads (Childers et al., 2000), while memory designers have improved energy usage by employing multi-level caches (L1 and L2 cache) for frequently accessed data-sets in memory (Kin et al., 1997).

The development of virtualization technologies allows IT managers to right-size infrastructure allocations for applications by consolidating applications on fewer servers at low workloads, and allocating increasing servers as workloads increase. This flexibility, supported by virtualization-aware request distribution strategies (Rajamani and Lefurgy, 2003), can result in significant energy savings (Chen et al., 2005).

The next logical step in the progression of this discussion is to ask about energy usage in the application layer. Indeed, there is some work in this area in the literature. Most of this research considers power usage in embedded software, which is designed for special-purpose scenarios such as pacemakers and sensor networks. Interest in energy efficiency is such use cases makes sense – the more efficient the software is, the longer the battery will last, resulting in longer intervals between replacements (Tiwari et al., 1994).

While there is some treatment of energy use in applications in the literature (e.g., Kansal and Zhao (2008) propose a fine-grained energy-use profiler for software), overall, the general-purpose application space is largely and conspicuously absent from the discussion of energy usage. Despite a long history of calls for leaner software (Wirth, 1995; Hyde, 2009), software applications continue to grow larger in functionality and size at a pace faster than Moore's Law can compensate for (Xu et al., 2010).

Hyde (2009) suggests that optimization, whether for performance or energy consumption, simply is not a priority in software development organizations. Hyde further hypothesizes that optimization in software development is hindered by a general lack of understanding of the implications of design-time choices, due to the fact that high-level programming languages obscure the details of the underlying implementations and the corresponding performance and energy use implications. Hyde suggests that improved optimization in the development process may not require significant additional time in the development schedule; rather, it may be a simple matter of education. If developers understand the performance and energy usage implications of the options within a set of equivalent functionality, they will automatically select the most efficient option.

Developers make thousands of small decisions every day, impacting everything from functional correctness to application performance to energy usage. While functional correctness and performance in software have received significant treatment in the literature and in industry, energy usage has not.

In this study, we *consider the energy usage impacts of some of these decisions*, focusing on some of the most basic operations, and *demonstrate that non-trivial energy savings can be achieved in software by making energy-conscious decisions regarding basic aspects of programming*. Our aim here is not to perform an exhaustive analysis across all programming constructs and programming languages; rather, we are simply proposing a first step toward greener software.

First, we focus on the energy impacts of implementation options for three primitive types of operations: (a) creating objects; (b) performing mathematical calculations; and (c) manipulating character strings. These operations form some of the fundamental building blocks of any application; at runtime, these operations may execute hundreds of thousands or millions of times per second (gigahertz processors are capable of executing billions of instructions per second).

Second, we consider the energy impacts for common, but less primitive, operations: (a) sorting items; (b) storing the results of computation to volatile or persistent storage; (c) supporting multiple simultaneous users through multi-threading; and (d) the effects of writing to persistent storage from multiple threads.

Our results show that significant energy savings can be obtained with some simple energy-conscious decision-making at software design time. These results have strong potential to impact practice: recent research shows that examples of positive results drive adoption of green IS and IT initiatives (Chen et al., 2009). A little knowledge can easily go a long way toward greener software – we simply need to build this knowledge base, and educate developers regarding the potential benefits of energy-conscious decision-making.

## 2.  Demonstrating the Energy Use Impacts of Design Time Decisions

In many respects, we can think of energy efficiency in software in terms of the familiar environmental tagline "reduce, reuse, recycle." We can *reduce* energy consumption by choosing the most efficient option among a set of equivalent functional options. We can *reuse* or *recycle* existing resources where possible to avoid the cost of creating new resources.

To demonstrate the potential for reducing, reusing, and recycling in application software, we developed a set of programs designed to test the software design choices noted above in isolation to observe the energy consumption impacts of each option. We coded each experiment in Java 1.6, and ran the experiments on a 64-bit Windows 7 server configured with a quad-core 2.2 GHz CPU and 8 GB RAM.

During each experiment, we recorded the server's energy consumption using the "Watts up? PRO" power meter (Watts Up?, 2011). With no user applications running, the server consumes 84.63 watts on average. In our experimental results, we reported the total energy consumption of the server for exclusively running each experiment. In our results, we report the total energy consumption of the server, including the baseline server energy consumption, rather than reporting the difference above the baseline for each experimental case.

In these experiments, operations run on randomly-generated values. Numeric values were generated using the Java Random class based on a uniform distribution as follows: the `float` and `double` range is 0.0 to 0.1, while the `int` range is $-2^{31}$ to $2^{31}$. Characters were randomly selected from the ASCII character set. Our results reported here do not include the energy cost of generating values. In each experiment, we first pre-create the appropriate number of values required, and then perform the operations of interest. We report energy consumption only for the experimental period when the operations of interest are running (the power meter's time is synchronized to the server's time via USB connection).

For each experimental case, we run the experiment 10 times, and report the average energy consumption in our results.

We now move on to describe our experimental results.

## 2.1 Comparing Energy Consumption for Object Creation and Object Reuse

In this experiment, we explore the energy consumption effects of reusing an instantiated object, rather than creating multiple instantiations. We created a Java class called `Simple` with a single `int` data member, a simple unparameterized constructor containing no logic, and a `set` method for the data member. We created two Java console applications, each containing a single loop. In the object creation case, a `Simple` object is instantiated within the loop, and its data member is set to an random integer value. In the object reuse case, the `Simple` object is instantiated prior to the loop; inside the loop, its data member is set to a randomly-generated integer.

| Task | Number of Objects | Duration (ms) | Joules (Watt-Sec) | Average % Energy Savings in Object Reuse Compared to Object Creation |
|---|---|---|---|---|
| Object Creation | 100,000 | 9 | 0.7497 | |
| | 1,000,000 | 30 | 2.517 | |
| | 10,000,000 | 226 | 18.9614 | |
| | 100,000,000 | 20582 | 2410.1522 | |
| Object Reuse | 100,000 | 6 | 0.4998 | 33.33 |
| | 1,000,000 | 10 | 0.839 | 66.67 |
| | 10,000,000 | 40 | 3.356 | 82.30 |
| | 100,000,000 | 993 | 118.5642 | 95.08 |

Table 1: Energy Consumption for Object Creation and Object Reuse

We report the average energy consumption for the creation and reuse cases in Table 1. In each experiment, we ran $10^I$ iterations of the loop, where $I = 5, 6, 7, 8$. Our results show that there is a significant energy consumption difference between the creation and reuse cases, with the reuse case showing 95% energy savings over the creation case for $I = 8$. Clearly, there is a strong case for *reusing* or *recycling* instantiated objects where possible to avoid the overhead associated with object creation.

## 2.2 Comparing Energy Consumption across Numeric Data Types

In this experiment, we consider energy consumption for mathematical operations across the four most frequently used data types in application programs: `int`, `float`, `double` and `boolean`. For each data type, we perform mathematic operations across $M$ pairs of data values. For `int`, `float` and `double` values, the operation is selected from the following operations: addition, subtraction, multiplication and division. For `boolean`, the operation is selected from the following operations: AND, OR, NOT and XOR.

We ran the experiment for two values of $M$, 100 million and 1 billion. We report the average time to run and energy consumption in Table 2.

Table 2 shows that there is considerable difference in energy consumption associated with the data types. Computations involving `int` require 12-13% less energy than `float`. Computations for `int`

| Number of Operations | Data Type | Time of Computation (milliseconds) | Average Power (Watt) | Energy Consumption (Joules) | % Energy Savings in Integer |
|---|---|---|---|---|---|
| 100,000,000 | int | 6074 | 85.7 | 520.5418 | 0 |
|  | float | 6473 | 91.34 | 591.24382 | 11.96 |
|  | double | 10426 | 91.77 | 956.79402 | 45.60 |
|  | boolean | 6066 | 91.56 | 555.40296 | 6.28 |
| 1,000,000,000 | int | 60610 | 85.81 | 5200.9441 | 0 |
|  | float | 64615 | 92.29 | 5963.31835 | 12.78 |
|  | double | 104157 | 92.37 | 9620.98209 | 45.94 |
|  | boolean | 60670 | 92.33 | 5601.6611 | 7.15 |

Table 2: Comparing Energy Consumption across Numeric Data Types

require 46% less energy than `double` data types.

Surprisingly, the computational overhead of `int` operations is 6.3% less than that of `boolean` operations, even though `boolean` is logically a one bit data type. Intuitively, it would seem more logical that comparing two single-bit values should require less energy than mathematical operations across 32-bit values. However this ignores that the fact that the `boolean` data type is more complex than it seems to be at first glance. This is because it is emulated by a 1 byte data type, which needs to ensure that only the lowest bit is set to 0 or 1, (false or true), and that the rest of the bits are set to always 0. This extra logic in the `boolean` data type is associated with additional overhead, which is responsible for the additional energy consumption in the `boolean` case as compared to the `int` scenario.

In application programs, a `boolean` data type can be replaced by an `int` with {0, 1} value. This is a common practice in many programming scenarios – the C language does not even have a standard boolean data type.

Our results show that considerable energy is wasted when the `double` data type is used and an `int` or `float` would suffice for the purpose, as well as when the `boolean` data type is used and the same functionality can be achieved with an `int`. These results represent an opportunity to *reduce* energy consumption in software.

| Task | Number of Objects | Duration (ms) | Joules (Watt-Sec) | Average % Energy Savings in StringBuffer Compared to String Creation |
| --- | --- | --- | --- | --- |
| String | 10,000 | 159 | 18.921 | |
| | 100,000 | 7789 | 926.891 | |
| | 1,000,000 | 1144689 | 136217.991 | |
| StringBuffer | 10,000 | 2 | 0.238 | 98.74 |
| | 100,000 | 8 | 0.952 | 99.89 |
| | 1,000,000 | 23 | 2.737 | 99.99 |

Table 3: Energy Consumption across String Representations

## 2.3 Comparing Energy Consumption across String Representations

In this experiment, we consider the energy consumption impacts of two different string representations: `String` and `StringBuffer`. In Java, a `String` is immutable; changing it requires creating a new `String` object. In contrast, `StringBuffer` allows changes to a sequence of characters. For these experiments, we created two Java console applications, each containing a single loop. In the string case, a `String` object is instantiated prior to the loop, and a single randomly-generated character is appended to the `String` in each loop iteration. In the string buffer case, a `StringBuffer` object is instantiated prior to the loop, and a single randomly-generated character is appended to the `StringBuffer` in each loop iteration. We ran this experiment for 10,000, 100,000 and 1 million loop iterations. Table 3 shows the results of these experiments.

The difference in energy consumption between `String` and `StringBuffer` data types is very significant – 98% less energy is consumed in the string buffer case across 10,000 append operations, as compared to energy consumption in the string case. This difference in energy use is due to the fact that `StringBuffer` effectively *reuses* or *recycles* a character string as it changes, while `String` creates a new object with each change.

While the `StringBuffer` data type is slightly more complex work with in code, the energy consumption overheads associated with the `String` data type more than outweigh the implementation convenience it offers. The only energy-conscious use case for `String` is the case where the character string

| Number of Integers | Algorithm | Time to Run (milliseconds) | Average Power (Watt) | Energy Consumption (Joules) | % Energy Savings compare to Qsort |
|---|---|---|---|---|---|
| 10,000,000 | Qsort | 1590 | 83.7 | 133.083 | 0 |
|  | Arrays.sort | 74 | 83.7 | 6.1938 | 95.35 |
|  | Collections.sort | 1175 | 93.4 | 109.745 | 17.54 |
| 50,000,000 | Qsort | 8635 | 94.036 | 812.0009 | 0 |
|  | Arrays.sort | 163 | 97.3 | 15.8599 | 98.05 |
|  | Collections.sort | 1498 | 112.85 | 169.0493 | 79.18 |
| 100,000,000 | Qsort | 17799 | 97.8 | 1740.742 | 0 |
|  | Arrays.sort | 276 | 98.3 | 27.1308 | 98.44 |
|  | Collections.sort | 3044 | 118.68 | 361.2619 | 79.25 |

Table 4: Energy Consumption across Sorting Methods

is known at object instantiation time, and will not change.

## 2.4 Comparing Energy Consumption across Sorting Methods

In these experiments, we compared the energy consumption impacts of three different types of sorting: (a) `Qsort`, an implementation of the quicksort algorithm (Hoare, 1962), which is known to give on average $O(n \ log \ n)$ complexity and is the most widely used sort algorithm in practice; (b) Java's `Array.sort`, which is a tuned quicksort, adapted from Bentley and McIlroy (1993); and (d) Java's `Collections.sort`, which is a modified mergesort where the merge is omitted if the highest element in the low sublist is less than the lowest element in the high sublist.

In each experiment, we apply each of the three sort options to sort $N$ integers. We ran the experiment for three values of $N$ (10 million, 50 million, and 100 million). We measured the total power usage in the computer during the sorting operation, and report our results in Table 4.

Based on Table 4, we can see that `Arrays.sort` provides the most efficient performance in terms of energy consumption, showing energy savings in the range of 95-98% compared to the generic `Qsort` algorithm. `Collections.sort` is more efficient than `Qsort`, but less efficient than `Arrays.sort`. For lower values of $N$, `Collections.sort` consumes 17% less energy than `Qsort`; for 50 million or above integers it uses 80% less energy than `Qsort`. The energy savings reported here is primarily due to reductions in the running time of the algorithm in the `Arrays.sort` and `Collections.sort` cases.

| Operations | Duration (milliseconds) | Data Read/Written MBytes | Average Power (Watt) | Energy (Joules) | % Energy Savings in Memory I/O |
|---|---|---|---|---|---|
| Disk I/O | 15,000 | 1,936 | 98.78 | 1481.7 | |
| | 30,000 | 3,870 | 99.3 | 2979 | |
| | 60,000 | 7,589 | 96.4 | 5784 | |
| Memory IO | 15,000 | 39,373 | 88.98 | 1334.7 | 9.92 |
| | 30,000 | 78,564 | 90.62 | 2718.6 | 8.74 |
| | 60,000 | 157,352 | 95.1 | 5706 | 1.35 |

Table 5: Comparing Energy Consumption for Disk and Memory Input/Output

While the `Qsort` algorithm is one of the most popular and widely used algorithms in practice, both `Collections.sort` and `Arrays.sort` are sort implementations easily available in one of the most widely used development frameworks, the Java Development Kit. Based on this experiment, we can see that choosing a sorting implementation in an energy conscious way can help *reduce* energy usage in an application.

## 2.5 Comparing Energy Consumption for Disk and Memory Input/Output

In this section, we analyze the energy consumption for disk input/output (I/O) as compared to memory I/O. A disk drive is typically an electromechanical device (solid state drives are available, but not in wide usage in data centers, so we do not consider this in our analysis), requiring mechanical movement (to spin the disk, move the reader head, etc.). In comparison, memory is solid-state, and the cost of writing to memory primarily involves the CPU processing required for memory management.

We randomly read and write for $T$ milliseconds from/to a disk using a single-threaded Java program. We then randomly read and write for the same duration from/to memory, which is a CPU-intensive operation. We ran the experiment for 3 different values of $T$ (15,000, 30,000 and 60,000). We measured the power and energy consumption for both the disk read-write and the memory read-write scenario. Our results are shown in Table 5.

As is clear from the results in Table 5, disk I/O requires significantly more energy than memory I/O. This is to be expected, since mechanical operations for the disk are more energy-expensive than memory

| Number of Threads | CPU Utilization (%) | Average Power (Watt) | % Power Used by Application | % Power Overhead for the Machine |
|---|---|---|---|---|
| 1 | Idle CPU | 84.63 | 0 | 100.00 |
| 1 | 25 | 93.21 | 9.20 | 90.79 |
| 2 | 50 | 102.31 | 17.28 | 82.72 |
| 3 | 75 | 110.89 | 23.68 | 76.32 |
| 4 | 100 | 119.15 | 28.97 | 71.03 |

Table 6: Energy Consumption across Levels of Multithreading

I/O, which primarily requires CPU. The total amount of work accomplished in the test period is also dramatically different in the two cases: the total bytes accessed in the disk case is only 5% of that accessed in the memory case in the same experiment length.

As the duration of the experiment increases, the percentage of additional energy consumed in the disk case decreases compared to the memory case. The initiation of disk I/O requires the movement of the head to a position where it can read or write. After this, the mechanical movement of the disk head is reduced significantly, resulting in lower energy consumption.

We consider two data points from Table 5. First, for longer duration read-write operations ($T = 60$ seconds) memory takes 1.3% less energy than disk read-write operations for the same duration. Second, the disk can read/write only about 5% of the data can be read/written with memory-based I/O operations within the same duration. Based on these observations, we estimate that disk I/O will require about 27 times more energy to read/write the same volume of data, as compared to memory based I/O.

## 2.6 Comparing Energy Consumption across Levels of Multithreading

In these experiments, we consider the energy consumption effects of supporting multiple simultaneous users through multithreading. We developed a multi-threaded program that does simple arithmetic computation and reads from and writes to memory. We vary the number of threads from 1 to 4 (one thread per CPU core), and report the results in Table 6.

When the server in our experimental testbed is idle (running only basic operating system tasks), the average power usage is 84.63 watts. This power usage is invariant of whether any application utilizes the computer; this power will be consumed to keep the machine running. As CPU utilization increases with the number of threads, the power consumption increases as well. Because the server is equipped with a quad-core CPU, at 4 threads our CPU-intensive program utilizes 100% of the available CPU cycles.

At full CPU utilization, the power consumption is about 40% higher than the idle state. With the increased number of threads, though the total CPU utilization and power consumption increase, one needs to note that the power usage when the server is idle is overhead – wasted energy if no computation is happening. This overhead portion of the power decreases significantly with the increased CPU utilization and the power consumption of the machines. At 4 threads, when the CPU utilization reaches 100% the overhead portion of power consumption is 71% of the total power consumption. In comparison, for a single-threaded application, which can only take advantage of a single CPU core for a maximum of 25% of CPU, the percentage of power use that is attributed to overhead is 91%.

A naive interpretation of the data might result in a conclusion that more threads leads to greater power utilization. Rather, a careful analysis of the data points reveals that higher number of threads simply amortizes the overhead of running the server (as a percentage of the total power consumption) across a larger number of jobs.

## 2.7 Comparing Energy Consumption across Levels of Multithreading in I/O-intensive Workloads

In these experiments, we consider how energy utilization changes with an increased number of threads in a disk-intensive application. We created an application with a varying thread count (from 1 to 4). Each thread in the application performs random disk read/write operations for 60 seconds. For each value of thread count, we report the average CPU utilization along with energy consumption in the Table 7.

It is quite common in application development to assume that multi-threading is always a good practice. From the Table 7, however, we observe that an increased thread count results in no significant change in energy consumption. We also note that an increased thread count results in a decreased volume of data accessed from the disk.

| Number of Threads | Disk IO (MBytes) | Duration (seconds) | Average Power (Watt) | Energy (Joules) | Energy/Data (Joules/MBytes) | % Savings in Energy/Byte in single-thread case |
|---|---|---|---|---|---|---|
| 1 | 7744 | 60 | 90.56 | 5433.6 | 0.7016 | 0 |
| 2 | 7380 | 60 | 89.78 | 5386.8 | 0.7299 | 3.87 |
| 3 | 7044 | 60 | 90.12 | 5407.2 | 0.7676 | 8.60 |
| 4 | 6364 | 60 | 89.56 | 5373.6 | 0.8443 | 16.90 |

Table 7: Energy Consumption across Levels of Multithreading in I/O-intensive Workloads

Based on a computation of the energy required to read/write 1 MB of data, we note that the energy/MB for data access actually increases with an increased thread count. For example, the 1-thread case requires 16.9% less energy than the 4-thread case to write 1 MB of data.

This counter-intuitive behavior can be explained only when one delves down into the details of how the disk works. The disk in the target computer is a single head ATA device. At any instant, the drive can execute only one read or write command. When the number of threads increases in an I/O intensive application, each thread dispatches I/O operations to the disk in parallel. However, the disk has to queue these requests and execute them sequentially. When the disk needs to execute commands from multiple threads, there is a switching cost incurred, which reduces the efficiency of I/O operations and increases the effective energy consumption per byte read or written.

Thus, in the above scenario, if an I/O-intensive multi-threaded application runs on the server, the CPU will remain underutilized and energy consumption will increase on a per-byte read/written basis. To make the system more energy efficient, we would need to install a RAID disk system, that can read and write using multiple heads in parallel. This would reduce the bottleneck in the I/O device and will improve the overall CPU utilization.

It is well known that higher power CPUs (with more cores and higher clock speed) are associated with higher energy consumption. It is customary to order computers with highest available processing power; however, if the disk is the bottleneck for an I/O intensive application, there will be a loss in energy efficiency.

# 3. Conclusion

This work represents a first step toward demonstrating that non-trivial energy savings is possible by focusing on software. We identify a few simple ways to improve energy efficiency in software by focusing on a few of the basic operations fundamental to any software component. Our results show that significant energy savings can be achieved with some basic knowledge of the effects of design-time decisions. These results can help guide developers' decision-making at software design time toward more energy-conscious design decisions to *reduce*, *reuse*, and *recycle*, resulting in greener software.